\documentclass{elsart}

\usepackage{graphics}
\usepackage{graphicx}
\usepackage{amssymb}
\usepackage{latexsym}

\begin{document}

\begin{frontmatter}



\title{Effect of kink-rounding barriers on step edge fluctuations}


\author{Jouni Kallunki, Joachim Krug}
\address{Fachbereich Physik, Universit\"at Essen, 45117 Essen, Germany}

\begin{abstract}
The effect that an additional energy barrier $E_{kr}$
for step adatoms moving
around kinks has on equilibrium step edge fluctuations is explored
using scaling arguments and kinetic Monte Carlo simulations. When 
mass transport is through step edge diffusion, the time correlation 
function of the step fluctuations behaves as $C(t) = A(T) t^{1/4}$. 
At low temperatures the prefactor $A(T)$ shows Arrhenius behavior with
an activation energy 
$(E_{det} + 3 \epsilon)/4$ if $E_{kr} < \epsilon$ and
$(E_{det} + E_{kr} + 2 \epsilon)/4$ if $E_{kr} > \epsilon$,
where $\epsilon$ is the kink energy and $E_{det}$ is the 
barrier for detachment of a step adatom from a kink. We point out
that the assumption of an Einstein relation for step edge diffusion
has lead to an incorrect interpretation of step fluctuation experiments,
and explain why such a relation does not hold. The theory is applied
to experimental results on Pt(111) and Cu(100).   

\end{abstract}

\end{frontmatter}

\section{Introduction}
In thin film growth the detailed knowledge of the microscopic elementary 
processes is essential since the large scale morphology is
determined by the competition between the nonequilibrium 
deposition flux and the
different relaxation mechanisms \cite{Zhang97}. 
Clear demonstrations of this  are 
for example growth instabilities on high symmetry or vicinal crystal surfaces,
which are known to produce self-organized nanoscale patterns 
\cite{Krug99,Politi00b}.
In both cases, mound formation on 
singular \cite{Villain91} and step meandering on
vicinal surfaces \cite{Bales90}, the size of the structures is set by
the relation between the time scales of deposition and 
relaxation kinetics \cite{Krug97a}.
Thus the knowledge of the relaxation kinetics opens a possibility to 
control the size of the structures by controlling the external parameters
such as deposition rate and temperature. 

The most important 
elementary process on surfaces is the hopping of individual atoms.
The motion of an adatom from one lattice site to another 
is a thermally activated process and takes place at rate
\begin{equation}
\Gamma_i = \Gamma_{i,0} \exp\left( - \beta E_i \right) ,
\end{equation} 
where $\beta=1 / k_B T$ is the inverse temperature, $\Gamma_{i,0}$ is an 
attempt frequency, 
and $E_i$ is the \emph{activation energy} of the process $i$.
Thus knowledge of the activation energies gives access to the elementary
rates.  
Unfortunately the activation energies are very rarely accessible
by direct measurement; it is extremely difficult to
follow a single atom diffusing on a surface and to extract
activation energies from its trajectory. Thus one has to rely
on measurements of mesoscopic quantities and try to compare these with
theoretical predictions in order to extract the microscopic parameters.
Examples of such an approach are the determination of the adatom
diffusion barrier from the island density \cite{Brune98}, 
the estimate of interlayer diffusion barriers from second layer
nucleation experiments and mound shapes
\cite{Krug00,Krug02}, and the extraction of activation energies for
processes at step edges from the characteristic
length scales of growth instabilities on vicinal surfaces
\cite{Maroutian99,Kallunki02,Rusanen01}.

An elegant method for measuring energetic and kinetic parameters
of atomistic processes at steps exploits the time correlation function
of equilibrium step fluctuations, see \cite{Jeong99,Giesen01}
for recent reviews.   
In this paper we revisit the theoretical basis of these
experiments for the case of mass transport dominated
by step edge diffusion, 
and take into account the possibility
of a significant \emph{kink rounding barrier}, which prevents
adatoms migrating along a step from hopping around a 
kink. 
The kink-rounding barrier is the one-dimensional analog of the well-known 
Ehrlich-Schwoebel (ES) barrier \cite{ES66} suppressing the inter-layer
mass transport on crystal surfaces. It is worth noting that
also a three-dimensional
analog of the ES barrier, inhibiting atoms going around
facet edges, has been observed \cite{Lagally02}.
The existence of the kink rounding barrier is still under debate;  
numerical calculations support its existence \cite{Merikoski97,Mehl99},
but until now experimental observations are few \cite{Buatier02,Kellog96}.
If present, the kink-rounding barriers have great impact on the
patterns formed under unstable epitaxial growth 
\cite{Kallunki02,Pierre-Louis99,Ramana99,Schinzer99,Amar99,Politi00a} 
as well as on the 
shape relaxation of islands and other
nanostructures \cite{Cadilhe00,Zhong01}. In this paper we show
how kink-rounding barriers affect equilibrium step fluctuations, 
thus providing an alternative way of determining the barriers
experimentally. In addition, in Sect.4 we clarify
a misconception in the theory of step fluctuations which has
lead to an incorrect data analysis in some cases.

\section{Step fluctuations and the adatom mobility}
 
On a vicinal surface the mono-atomic steps, separating high symmetry
terraces, are not perfectly straight but wander due to thermal 
fluctuations. Here we will consider a situation
where adatoms cannot detach from
the steps, thus the only possible mass transfer process is migration of 
adatoms along the steps. The Langevin theory of step
fluctuations then yields the expression \cite{Jeong99,Giesen01}
\begin{equation}\label{def_cor}
C(t) \equiv \langle \left[ \zeta(x,t) - \zeta(x,0)\right]^2 \rangle =
a_\perp^2 \frac{\Gamma(3/4)}{\tilde{\gamma} \pi \beta
\Omega^{1/2}} \left( 2 \sigma 
\tilde{\gamma} t\right)^{1/4} 
\end{equation}
for the time correlation function of the step edge position $\zeta(x,t)$
with the initial condition of a flat step,
$\zeta(x,0)\equiv0$.
Here $\tilde \gamma$ is the step stiffness, $a_\perp$ the lattice
spacing perpendicular to the step, $\Omega$ is the atomic area,
$\Gamma(3/4) \approx 1.2254...$, 
and $\sigma$ denotes
the adatom mobility along the step edge. It is
defined through the relation
\begin{equation}
\label{sigmast} j = - \sigma
\partial_x \mu = \sigma \partial_x \Omega \tilde \gamma 
\partial_{xx} \zeta
\end{equation}
between the mass current along the step and the chemical
potential gradient driving it. In the last equality
the Gibbs-Thomson relation has been used.

\begin{figure}
\begin{center}
\includegraphics[width=12cm,height=6cm,]{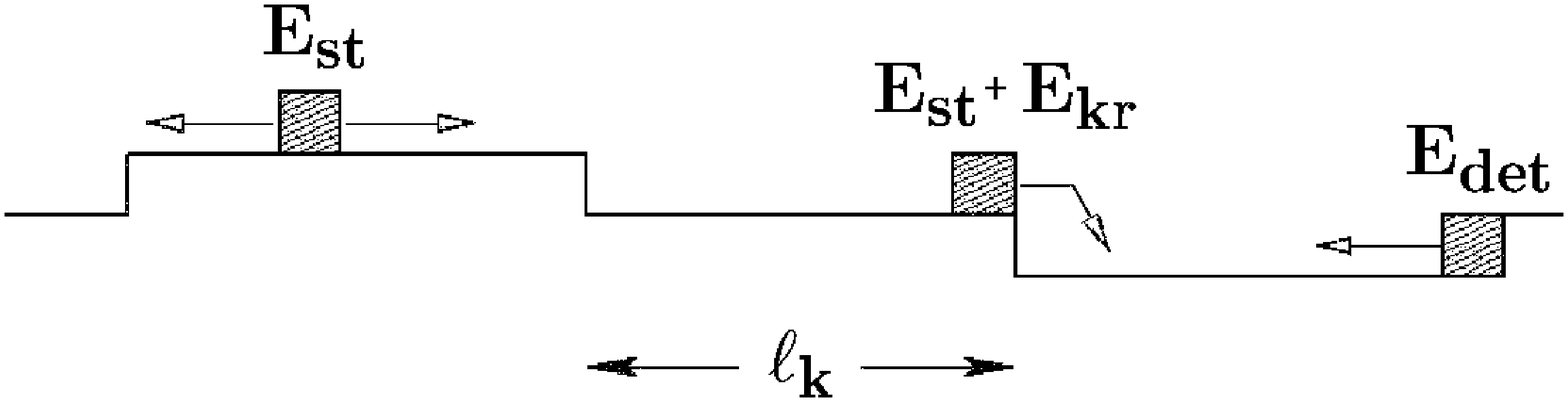}
\caption{A schematic picture of a monoatomic step with
the activation energies of the elementary processes.
}
\label{fig_step}
\end{center}
\end{figure}

To make use of the expression (\ref{def_cor}) for the analysis of
experimental data, 
the parameters $\tilde \gamma$ and 
$\sigma$ of the continuum description must be expressed in 
terms of the rates of the elementary processes.
This can be done exactly if the step is modeled as a one-dimensional
solid-on-solid (SOS) surface with energy barriers proportional to the
number of lateral bonds in the initial state 
(Arrhenius kinetics); see Sect.3 for 
a precise definition. In this case one obtains \cite{Krug95}
\begin{equation}
\label{sigmaSOS}
\sigma = \frac{a \Gamma_0 \beta}{2}\exp\left(-\beta E_{det}  \right) 
\end{equation}
and
\begin{equation}
\label{stiff}
\beta \tilde{\gamma} = a^{-1}(\cosh \left( \beta \epsilon \right) - 1), 
\end{equation}
where $E_{det}$ denotes the energy barrier for detachment of a step
atom from a kink site, the \emph{kink energy}
$\epsilon$ is the energy cost of creating a kink, $a$ is the
lattice constant parallel to the step, 
and $\Gamma_0$ is the attempt frequency, which
is assumed to be the same for all processes.
Since the detachment of a kink atom creates two kinks, within the
Arrhenius model $E_{det} = E_{st} + 2 \epsilon$, where
$E_{st}$ is the energy barrier for diffusion along the straight
(unkinked) step. Figure \ref{fig_step} shows 
\noindent
a cartoon of a step displaying the relevant processes.
The relation $\sigma \sim e^{-\beta E_{det}}$ has also been derived
within a Kubo formalism \cite{Pimpinelli93,Villain95}. 

Putting everything together yields
\begin{equation}
\label{C2}
C(t)/a_\perp^2 = g \times \left( \frac{t}{\tau_{st}}\right)^{1/4} 
\end{equation}
where $g$ is a numerical constant of order unity,
and $\tau_{st} \sim \tilde{\gamma}^{3}/\sigma$
is the characteristic time for the step to fluctuate one lattice
constant. In the low temperature limit $\beta \epsilon \gg 1$ one finds
\begin{equation}\label{t_char1}
\tau_{st} \sim \exp \left[ \beta (E_{det} + 3 \epsilon) \right].
\end{equation}
Equations (\ref{C2}) and (\ref{t_char1}) form the basis of 
the experimental determination of kinetic barriers from step edge
fluctuations. Measuring the coefficient of the $t^{1/4}$-behavior
of the correlation function (\ref{C2}), the
activation energy of the characteristic time (\ref{t_char1}) 
can be obtained.
Provided the kink energy $\epsilon$ is known from other sources 
(e.g., from the analysis of \emph{static} step fluctuations
\cite{Jeong99,Giesen01}), this yields an estimate of the detachment
barrier $E_{det}$.

To see how (\ref{t_char1}) has to be modified in the presence
of kink rounding barriers, we first rederive $\tau_{st}$
from a scaling argument.
The elementary process driving the step fluctuations is the transport
of an atom from one kink to another, which allows the kinks to diffuse
along the step.
In order to move the step by a distance $a_\perp$, a kink must
diffuse over a distance of the order of the mean kink
spacing $\ell_{k} = (1/2) a e^{\beta \epsilon}$.
The detachment rate of atoms from a kink site is
$\Gamma_{det} = \Gamma_{0}\exp(-\beta E_{det})$. 
The probability $P_{att}$ for an emitted
adatom to reach another kink at distance $\ell_{k}$ 
before it reattaches to the original kink 
can be calculated from
a random walk theory \cite{Kallabis97}, yielding 
$P_{att} \approx \ell_{k}^{-1}$. 
Thus
the diffusion rate of a kink is $\Gamma_{det}P_{att}$ and
the characteristic time for a step to fluctuate over a single lattice 
constant reads
$
\tau_{st} \sim \ell_{k}^{2}/(\Gamma_{det}P_{att}) 
\sim \exp[\beta (3 \epsilon + E_{det})]$,
in agreement with (\ref{t_char1}).

When atoms diffusing along the step experience an extra barrier $E_{kr}$
for going around a kink site, as drawn in Fig.\ref{fig_step},
the probability $P_{att}$ of the adatom to attach to a kink
at distance $\ell_{k}$ is altered. After reaching the distant kink
the adatom still has to overcome the kink rounding barrier in order to attach
to it. This yields \cite{Kallabis97} 
\begin{equation}
\label{Patt}
P_{att} \approx (\ell_{k}+1/p_{kr})^{-1},
\end{equation}
where
$p_{kr}\approx \exp(-\beta E_{kr})$ is
the probability for going around a kink.
Comparing $\ell_{k}$ with $p_{kr}^{-1}$ it is obvious that the
kink rounding barriers are relevant if $E_{kr} > \epsilon$.
In this case (\ref{t_char1}) has to be replaced by 
\begin{equation}\label{t_char3}
\tau_{st}\sim\exp\left[ \beta(2 \epsilon +E_{det}+E_{kr}) \right].
\end{equation}
As shown before the characteristic time is generally
a combination of the adatom mobility 
and the step stiffness, $\tau_{st} \sim
\tilde{\gamma}^{3}/\sigma$. Since the step stiffness
does not depend on the dynamics of the adatoms along the step, but only on the
energetics, we may conclude that the adatom mobility is reduced to
\begin{equation}\label{sigmareduc}
\sigma \sim \exp(- \beta(E_{det}+E_{kr}-\epsilon))
\end{equation} 
by the kink-rounding barrier, when $E_{kr}>\epsilon$. 
The expression (\ref{Patt}) suggests the interpolation
formula
\begin{equation}
\label{sigmagen}
\sigma = \frac{1}{2} a \Gamma_0 
\frac{e^{-\beta E_{det}}}{1 + e^{\beta (E_{kr} - \epsilon)}}
\end{equation}
for the mobility, which recovers (\ref{sigmaSOS}) for
$E_{kr} \ll \epsilon$. 

\section{Monte Carlo simulations}
\label{KMC}

In order to confirm the validity of the arguments of the previous section
we have conducted Monte Carlo simulations of a simple one-dimensional 
SOS model.
The position of the step at site $i$ is $h_{i}$ and the atoms 
may hop along the step to neighboring sites
($i \rightarrow i\pm 1$) with rate
\begin{equation}\label{sim_rates}
\Gamma_{i, i \pm 1} = \Gamma_0 \exp( -\beta E_{i, i \pm1} ) 
\end{equation}
\begin{figure}
\begin{center}
\includegraphics[width=10cm,height=7.5cm,]{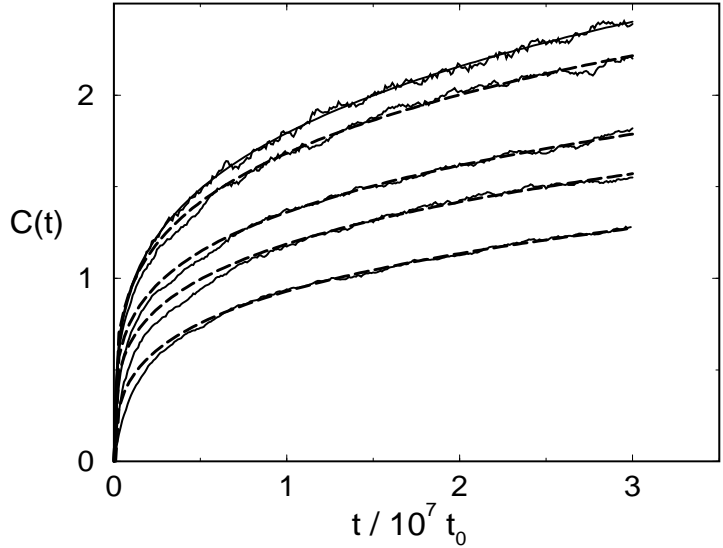}
\caption{The time correlation function (Eq. (\ref{sim_cor}))
for $\epsilon \beta=2.5$
and $E_{kr}/\epsilon = 0.0 / 0.4 / 1.2 / 1.6 / 2.0 $ (from top 
to bottom). The dashed line is the best fit $At^{1/4}+B$.
Time is measured in units of the inverse diffusion rate
along flat a step, $t_0 \equiv 1/(\Gamma_0 \exp[-\beta  E_{st}]) $
}
\label{fig_cor}
\end{center}
\end{figure}
\noindent
where the activation energy depends on the local configuration as
\begin{equation}
E_{i, i \pm 1} = E_{st} + 2 \epsilon n_{i} 
+ \left[ 1-\delta(h_{i}-h_{i \pm 1} -1) \right] E_{kr}. 
\end{equation}
Here $n_{i} = 0, 1, 2$ is the number of lateral 
nearest neighbors of the atom at initial site $i$ and $E_{kr}$ 
is an extra barrier suppressing kink rounding;
whenever the hop from $i\rightarrow i \pm 1$ 
is \emph{not} along flat step \emph{i.e.} $h_{i}-h_{i \pm 1} \neq 1$, 
the extra barrier $E_{kr}$ is added.
In the simulations we used a lattice of size $L=131072$, starting with a
straight step $h_{i}(0) \equiv 0$. 

The hopping rate $\Gamma_0 \exp[-\beta E_{st}]$ 
of a free step edge atom on an unkinked step
segment determines the time scale of the model,
and can be set to unity in the simulation. 
For the kink energy we used the value 
$\epsilon = 0.1$ eV. 
The kink-rounding barrier $E_{kr}$ was varied between 0 and 0.24 eV,
and the inverse temperature in the interval 
$\beta \epsilon = 1.25 - 3.5$, corresponding to $T = 331 - 928$ K.

The time-correlation function 
\begin{equation}\label{sim_cor}
C(t) \equiv \sum_{i=1}^{L} h_{i}(t)^{2} 
\end{equation}
measured from the simulations is shown in Fig. \ref{fig_cor}.
It has a clear $t^{1/4}$ time dependence and
the prefactor in Eq. (\ref{def_cor}) was determined by 
fitting the data in the long time limit with $C(t) = At^{1/4}+B$.
Using Eqs.(\ref{def_cor}) and (\ref{stiff}), 
the
\begin{figure}
\begin{center}
\includegraphics[width=10cm,height=7.5cm,]{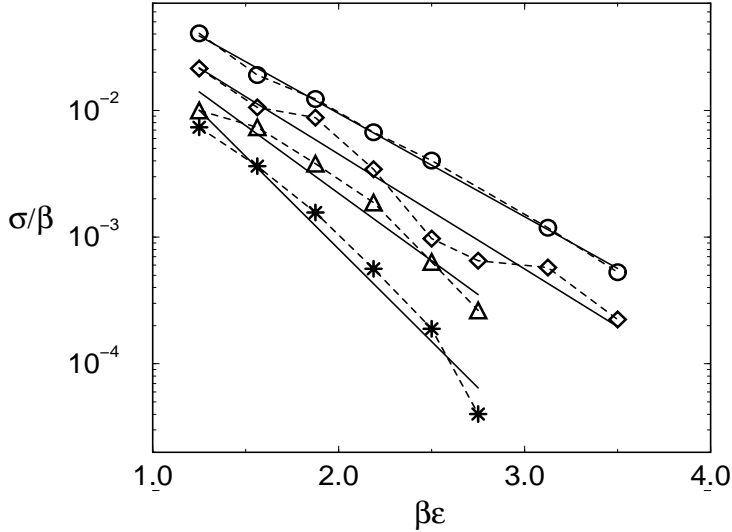}
\caption{The adatom mobility along the step edge 
obtained from the prefactor of the correlation function
in Fig. \ref{fig_cor}, with
$\epsilon = 0.1$ eV and kink rounding barrier 
$E_{kr}/\epsilon=0.0 (\bigcirc),  1.2 (\Diamond), 
1.6 (\bigtriangleup), 2.4 (\ast)$.
The full lines are best fits to an Arrhenius form.
}
\label{fig_mobility}
\end{center}
\end{figure}
\noindent
mobility $\sigma$ can be extracted from the
prefactor.  The mobility obtained from the simulation
results is shown in Fig. \ref{fig_mobility} and the activation
energy for the mobility $E_{\sigma}$, determined from
a fit to the Arrhenius plot, is shown in Fig. \ref{fig_E-sigma}. 
There is a cross-over in the behavior of the activation
energy $E_{\sigma}$ as the kink-rounding barrier roughly equals the kink 
energy $E_{kr} \approx \epsilon$. Thus the simulation results are in
good agreement with the analytical results (Eqs. (\ref{sigmaSOS}) and 
(\ref{sigmareduc})) of the previous section.

\section{No Einstein relation for step edge diffusion} 
\label{Einstein}

In the literature 
\cite{Giesen01,Giesen95,Giesen96} the
interpretation of experimental step fluctuation data is often
based on an Einstein relation \cite{Pimpinelli93b,Khare95,Khare98}
\begin{equation}
\label{Einstein_rel}
 \sigma = \frac{n_{st} D_{st}}{k_{B} T}
\end{equation}
for the mobility, where $n_{st}$ is the (one-dimensional)
concentration of step adatoms and $D_{st}$
denotes the tracer diffusion coefficient for an adatom migrating
along a kinked step. The latter can be estimated by considering the
motion of an adatom in a model potential where kink sites are
represented as traps of depth $E_{det}$ spaced at the mean
kink distance \cite{Giesen95,Giesen96}. 
The resulting activation energy for $D_{st}$
is $E_{det} - \epsilon$. Since a step adatom can be viewed
as a double kink, the concentration of
step adatoms in equilibrium is $n_0 \sim e^{-2 \epsilon/k_{B}T}$,
\begin{figure}
\begin{center}
\includegraphics[width=10cm,height=7.5cm,]{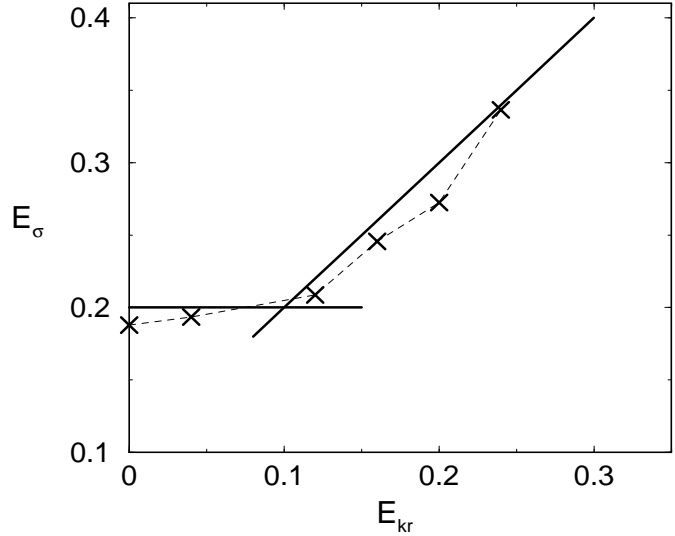}
\caption{The activation energy for the adatom mobility obtained
from the fits in Fig \ref{fig_mobility}.
A clear cross-over in the behavior
is seen at $E_{kr} \approx \epsilon=0.1$. The full lines are
the theoretical predictions Eqs. (\ref{sigmaSOS}) and 
(\ref{sigmareduc}).
}
\label{fig_E-sigma}
\end{center}
\end{figure}
\noindent
and hence the activation energy of $\sigma$ is
predicted by (\ref{Einstein_rel}) to be $E_{det} + \epsilon$,
in disagreement with the exact result (\ref{sigmaSOS}).

The problem with (\ref{Einstein_rel}) was already noted
in \cite{Villain95}. Since step adatoms are continually 
absorbed and emitted at kinks, they do not constitute a conserved
species, and it is difficult to consistently define
$D_{st}$ and $n_{st}$. 
In fact, strictly
speaking $D_{st} \equiv 0$:
It can be shown \cite{Krug96,Krug97b} that
the mean square displacement of a marked step adatom 
moving along a kinked step grows
\emph{sublinearly}, 
as $\langle (x(t) - x(0))^2 \rangle \sim t^{7/8}$. 
This reflects the fact that a trapped adatom runs a
considerable risk of being overgrown by a large step fluctuation,
and hence the probability distribution of trapping times has a
very broad tail. Representing the
migration of an adatom along a kinked step by the diffusion of a
particle in an external potential neglects both the possibility of
long term trapping due to step fluctuations, and the fact that a
kink site is not actually ``filled'' when an adatom attaches to it
-- it is merely shifted. 

Nevertheless a relation of the form
(\ref{Einstein}) does hold, if $D_{st}$ is replaced by the
diffusion coefficient $(1/2) a^2 \Gamma_0 \exp[-\beta E_{st}]$ 
for the migration along a
\emph{straight} close packed step, and the bond counting
relation $E_{det} = E_{st} + 2 \epsilon$ is assumed. 
Then (\ref{Einstein_rel}) 
simply expresses the balance between the detachment and attachment
of step adatoms at the kinks.

\section{Conclusions}

We have studied the time fluctuations of a monoatomic step when
the mass transport is restricted to migration along the step. The scaling
arguments presented in this paper [Eqs. (\ref{t_char3}) and (\ref{sigmareduc})],
show how the adatom mobility along
the step is reduced if the kink-rounding hops are suppressed with an extra
barrier. The results of our Monte-Carlo simulations confirm the validity of
the scaling arguments. 

Time-dependent step fluctuations open a possibility
for the experimental measurement of the activation energies
of elementary processes on a stepped surface \cite{Jeong99,Giesen01}.
The results presented here 
show that kink-rounding barriers 
have to be taken into account in the analysis of such experiments.
Provided the kink energy $\epsilon$ is known, the temperature
dependence of the prefactor of the time correlation function $C(t)$
gives access to the activation energy $E_\sigma$ of the adatom mobility.
In general, Eqs.(\ref{sigmaSOS}) and (\ref{sigmareduc}) show 
that $E_\sigma$ is an upper bound on the detachment barrier $E_{det}$.
For example, the interpretation of the 
results presented in \cite{Giesen96} in the light of our
work shows that $E_{det} \leq 
1.50 \pm 0.16$ eV for close-packed steps on 
Pt(111).   

When additional information on the energy barriers is available, 
the analysis of time fluctuations may be used to determine
the strength of the kink-rounding barrier itself. 
For the close-packed [110]-step on Cu(100), the re-interpretation of
the experimental
results of \cite{Giesen95} for the time fluctuations, 
supplemented with a result of \cite{Maroutian99} for
the diffusion barrier $E_{st}$ at a straight step, 
yields the estimate $E_{kr}\approx 0.41$ eV \cite{Kallunki02}. 
 
\section*{Acknowledgements}

Useful discussions with H.J. Ernst, M. Giesen, M. Rusanen, S.V. Khare and T. Michely
are gratefully acknowledged. This work was supported by DFG within
SFB 237. 



\end{document}